# Simulation of the reaction of fragmentation of deuterons into cumulative and twice-cumulative pions

A.G. Litvinenko[1,2], E.I. Litvinenko[3]
1. JINR, VBLHP, Dubna, Moscow region.
2. Dubna International University, Dubna, Moscow region.
3. JINR, FLNP, Dubna, Moscow region.

## Abstract

The paper discusses the features of the behavior of pion production cross section as a function of the atomic mass of the target nucleus for the reaction of fragmentation of the incident deuterons in the pions produced in the so-called "twice-cumulative" kinematic region. A distinctive feature of the twice-cumulative pions is that for their production the target nucleus $A_t$ must be heavier than hydrogen. The simulation results show that the dependence of the cross sections for pion production in the twice-cumulative region differs from the analogous dependence for the cumulative region. The paper discusses the reasons for such differences. In this work we present simulation results for some models of nuclear structure on small nucleon-nucleon distances.

## Definitions

To clarify the terminology and purpose of the work we will focus briefly on the definitions that will be used in this work (see, for example, [1-5]). To do this, consider the inclusive production of the peony in the experiment fixed target:

$$A_b + A_t = \pi(\theta) + X \ . \tag{1}$$

Here $A_b$ and $A_t$ are the beam and the target nuclear, respectively, $\theta$ is the angle between the momentum of the pion production and the direction of the beam, and the energy per nucleon is assumed fixed. The maximum energy of the pion, which it may have in the relevant reaction is denoted $E_\pi^{max}(A_b + A_t \to \pi(\Theta))$. This value depends from the angle of pion production, the initial energy and the atomic masses of the colliding nuclei. In the future, we will distinguish pions produced in the target fragmentation region $\Theta > 90^O$ and in the beam fragmentation region $\Theta < 90^O$. If the energy of the pion produced in the backward hemisphere (in the target fragmentation region) is greater than the energy available in the collision of a proton with a hydrogen target, such pion called cumulative produced in the fragmentation of the target. Consequently, the cumulative energy of the pion, born in the fragmentation of the target must satisfy the inequality

$$E_\pi > E_\pi^{max}(p+p \to \pi(\Theta)); \ \ \Theta > 90^O \ . \tag{2}$$

By analogy, the area for the kinematic variables is determined for the cumulative pions produced in the fragmentation region of the beam nucleus:

$$E_\pi > E_\pi^{max}(p+p \to \pi(\Theta)); \ \ \Theta < 90^O \ . \tag{3}$$

From these definitions, it follows that at the production of cumulative pion in the backward hemisphere the target nucleus must be heavier than the proton $A_t > 1$, and for the production of cumulative pion in the forward hemisphere the beam particle must be heavier than the proton $A_t > 1$

To clarify the concept of "double-cumulative pion", which is used in this paper (see [5]) consider the following inclusive reactions:

$$p + p = \pi(0^o) + X. \qquad (4)$$

$$D + p = \pi(0^o) + X. \qquad (5)$$

Pions with energy greater than permitted in the collision of incident deuterons with protons will be called twice-cumulative. This means that for the energy of the twice-cumulative protons the following inequality is performed:

$$E_\pi \geq E_\pi^{max}(D + p \to \pi). \qquad (6)$$

This means that the production of the twice-cumulative pions is possible only in the reaction of collision of the incident deuteron with the target nucleus heavier than hydrogen nucleus, i.e. in the reaction:

$$D + A_t = \pi(0^o) + X; \ A_t > 1. \qquad (7)$$

Dependence on the initial energy of the maximum values of the energy of pions, produced at zero angle, for different combinations of the target nuclei and the beam nuclei are shown in Fig. 1.

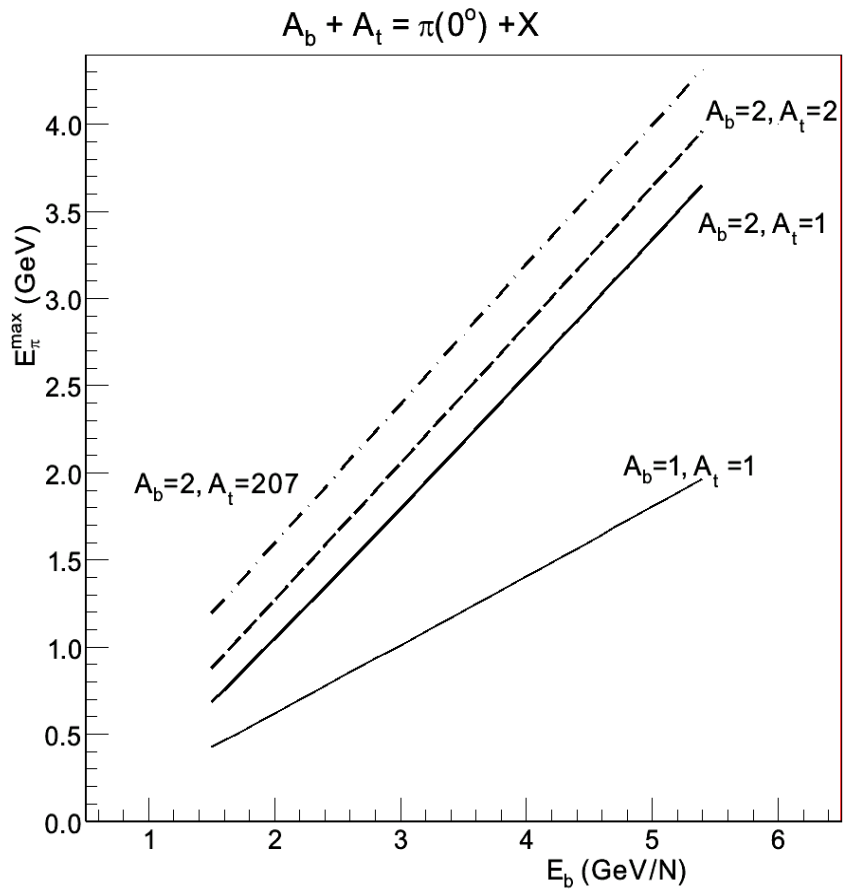

Fig. 1. Dependence of the maximum energy of pions on the initial energy per nucleon for different combinations of the colliding target nuclei and beam nuclei.

According to the above definitions, the energies of the cumulative pions lies between the curves marked as $A_b = 1, A_t = 1$ and $A_b = 2, A_t = 1$. The energy of the twice-cumulative pions lies in the region above the curve marked as $A_b = 2, A_t = 1$.

## Motivation

The colliding nuclei are included in the definition of cumulative particles asymmetrically. In the experiment, this asymmetry leads to a different dependence of the production cross section on the atomic weight of the target nucleus for the beam particle fragmentation [5] and for the target nucleus fragmentation [6]. The corresponding experimental data are shown in Fig. 2 and Fig. 3. From these data it is clear that in the case of fragmentation of the incident deuterium into cumulative pions the dependence of the cross section on the atomic mass of the target nucleus for medium and heavy nuclei $A_t > 12$ is close to the peripheral one $d\sigma \propto A_t^{0.4}$ [5]. And in the case of fragmentation of the target nucleus in the cumulative pions the dependence of the cross section on the atomic mass of the target nucleus for medium and heavy nuclei $A_t > 12$ has the bulk nature $d\sigma \propto A_t^{1.1}$.

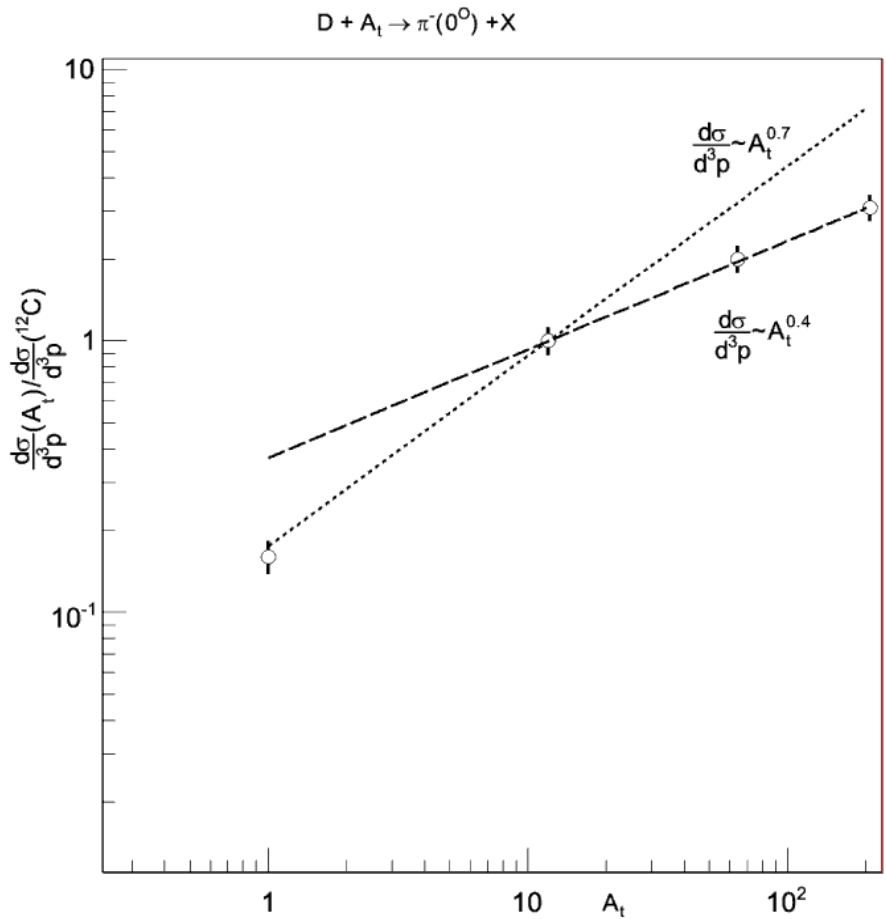

Fig. 2. Dependence of the cross section of cumulative pions production on the atomic mass of the target nucleus in the fragmentation of the beam nucleus (data from [5]). The cross section is normalized to the cross section of a carbon target. Lines correspond to the appropriate degree of dependence.

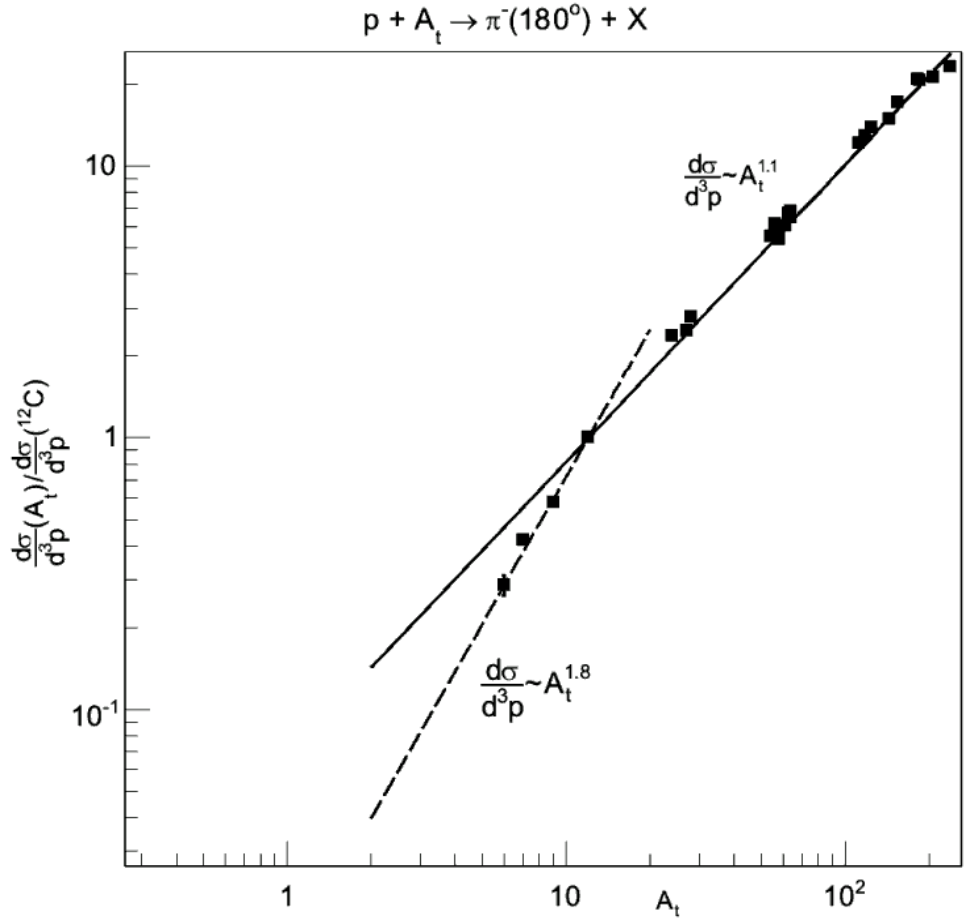

Fig. 3. Dependence of the cross section of cumulative pions production on the atomic mass of the target nucleus in the fragmentation of the target nucleus (data from [6]).The cross section is normalized to the cross section of a carbon target. Lines correspond to the appropriate degree of dependence.

In the conventional to the present time models with cold flucton cumulative particles are produced due to the high momentum components in the fragmenting nucleus [1-4]. The source of such high momentum components is configurations in which two or more nucleons are separated by distances smaller than the average distance between the nucleons in the nucleus [1-4]. According to [7], such configuration is called flucton. In models with a cold flucton cumulative pion production in the beam fragmentation region and in the target fragmentation region is different in that in the first case the flucton is needed to be in the beam nucleus, and in the second case the flucton should be in the target nucleus. It is this difference determined in the behavior of the dependence on the atomic mass of the target nucleus for the cross section of a cumulative pion production in the beam fragmentation region (Fig.2) and in the target fragmentation region (Fig.3). The concrete structure of a flucton is not important for us for further consideration, i. e. it does not matter whether we describe the properties of a flucton based on the nucleon degrees of freedom (as proposed in [8]) or the nucleons in a flucton overlap (and then the nonnucleon degrees of freedom must be taken into account at the description of the production of cumulative particles [1,2]).

At the deuteron fragmentation (reaction (4)) the cumulative pion is produced in the incident deuteron collision with one of the nucleons of the target nucleus. This production occurs due to high-momentum component of the wave function of the deuteron. I.e. when passing through the target nucleus the incident deuteron should not experience inelastic interactions until the production of the cumulative pion. Because of the rapid decrease of the cross section of cumulative pions with increasing momentum and emission angle it is necessary for the produced pion to leave the nucleus of the target without collision (for details see [9]). This means that the cross section is defined by the following expression:

$$d\sigma/d^3p \propto \iint dz db b (\sigma(NN \to \pi) n_N(z,b)) \overline{W}_D([-\infty,z],b) \overline{W}_\pi([z,\infty],b) , \qquad (8)$$

where b and z are the impact parameter and the value of the coordinate along the trajectory of the deuteron, $\overline{W}_D([-\infty,z],b)$ denotes the probability for the deuteron to reach the point of production of cumulative pion without interaction, $\overline{W}_\pi([z,\infty],b)$ is the probability for the produced pion to leave the target nucleus without scattering, and $n_N(z,b)$ is the nucleon density at the point of the cumulative pion production.

According to such a space-time picture, the main contribution to the cross section comes from the trajectories the total length of which is close to the sum of the mean free path of the deuteron and pion. The mean free path of the deuteron $\lambda_D \cong 1$ fm and the pion $\lambda_\pi \cong 2.4$ fm for the middle and heavy nucleus. This means that for medium and heavy nuclei the main contribution to the cross section comes from the large impact parameters (see [9]). This dependence is confirmed by experimental data on the dependence of the cross section of deuteron fragmentation into cumulative pions [5, 11-12]. The obtained cross section for the production of cumulative pions in these articles can be approximated by the following dependence:

$$d\sigma/d^3p(D+A_t \to \pi + X) \propto A_t^\alpha; (A_t \geq 12, \alpha \cong 0.4) . \qquad (9)$$

It is noteworthy that the similar dependence in the non-cumulative region is close to the surface dependence $\alpha \cong 0.6$ [11]. In the framework of this mechanism such kind of dependence is distinct due to the fact that for pion production in the non-cumulative field it is not necessary to require the pion to be produced by deuteron as a whole, and that the pion is to leave the target nucleus without interaction.

In turn, in the considered space-time picture it is necessary to take into account that twice-cumulative pions can be produced only in the scattering of the deuteron on the flucton in the target nucleus. This means that there should be the density of the fluctons $n_F(z,b)$ in the expression for the cross section unlike the case of the cumulative pions production:

$$d\sigma/d^3p \propto \iint dz db b (\sigma(NN \to \pi) n_F(z,b)) \overline{W}_D([-\infty,z],b) \overline{W}_\pi([z,\infty],b) . \qquad (10)$$

Note that this expression differs from the expression for the cross section of cumulative pions (8) by replacing in the integral of the density of the nucleons by the density of the fluctons in the target nucleus $n_N(z,b) \to n_F(z,b)$ .

In this paper, it is shown that the dependence of the cross section of twice-cumulative pions on the atomic mass of the target nucleus is sensitive to the distribution of fluctons, and therefore the experimental data on the cross section of twice-cumulative pions allow one to critically evaluate different models proposed to describe the distribution of fluctons by the volume of the nucleus. The results of the simulation of the cross section production of twice-cumulative pions (reaction (7) under the condition (6)) for the two models of the spherical flucton and for the cylindrical flucton (tube model) are discussed.

## Simulation

In the simulation of one event in the first stage of the collision the position of the nucleons in the colliding nuclei is determined. For medium and heavy nuclei the Wood-Saxon distribution [13] was used:

$$P_A(r) = \frac{N}{1 + \exp((r - R_A)/d)}, \quad (11)$$

where $r$ is the distance to the center of the nucleus, and $d$ is the diffuseness parameter. The nuclear radius was chosen to be equal to:

$$R_A = r_0 \cdot A^{1/3}, \quad (12)$$

with $r_0 = 1.2$ fm. The normalization constant $N$ was chosen from the condition:

$$\int_0^\infty P_A(r) r^2 dr = 1. \quad (13)$$

The distribution of distances between the proton and the neutron in deuteron was chosen in accordance with the Hulthen wave function [14]:

$$P_D(r_{pn}) = \frac{2ab(a+b)}{(a-b)^2} \cdot \frac{(\exp(-2ar_{np}) + \exp(-2ar_{np}) - 2\exp(-(a+b)r_{np})}{r_{np}^2}, \quad (14)$$

where $r_{np}$ is the distance between the proton and the neutron, $a = 0.228$ (1/fm), and $b = 1.7$ (1/fm). The position of the nucleons in $^4He$ was described by the normal distribution [15]:

$$P_{^4He}(r) = \frac{4}{\sqrt{\pi} d^3} \cdot \exp(-r^2/d^2), \quad (15)$$

with $d = 1.7$ fm.

The center of the target nucleus was chosen at the origin. After that the coordinates of all the nucleons of the target nucleus and beam nucleus were independently played out. At the same time, this state is in accordance with one of the distribution (12), (13) or (14). The Z axis is directed along the beam. The impact parameter b was uniformly chosen inside the circle of the radius $R_b$:

$$R_b = \sqrt{x^2 + y^2} = 1.2 \cdot (R_{A_b} + R_{A_t} + 1.2), \quad (16)$$

where $R_{A_t}$ and $R_{A_b}$ are the radii of the target nucleus and the nucleus of the beam, respectively. For the nuclei with $A \geq 12$ the nuclear radius was calculated in accordance with (12), $R_D = 5$ fm was taken for the deuteron, and for $^4He$ $R_{^4He} = 3$ fm considered. After fixing the coordinates of the nucleons in the colliding nuclei the scattering of nucleons was considered by the following scheme. Along the path of the hadron the cylinder with the radius $r_{hN} = \sqrt{\sigma_{hN}/\pi}$ was chosen. If this cylinder got the nucleon from the target nucleus, the selected hadron was believed to have experienced the collision. If the cylinder got several nucleons, the first in the time collision was taken into account. Likewise, the interaction of secondary hadrons was described, but the speed of the secondary hadron was not directed along the axis of the collision. The following values of the total cross sections were used in further calculations:

1. nucleon – nucleon scattering

$$\sigma_{NN} = 45\,\text{мб} \Rightarrow r_{NN} = 1.197\,\text{фм}, \qquad (17)$$

2. pion – nucleon scattering

$$\sigma_{\pi N} = 30\,\text{мб} \Rightarrow r_{\pi N} = 0.977\,\text{фм}. \qquad (18)$$

The coordinates of the cumulative particle were selected equal to the coordinates of the nucleon in deuteron, which collided first with the nucleon of the target nucleus.

Let us consider three possible mechanisms of the production of the twice-cumulative pions in the deuteron fragmentation.

1. The direct mechanism in which pions are produced in the collision of one of the nucleons of the incident deuteron on flucton in the target nucleus, and the produced pion leaves the target nucleus without interaction:

$$D(pn) + A_t = \pi(0^o) + X. \qquad (19)$$

2. The cascade mechanism with an intermediate pion. In this mechanism the intermediate pion $\pi_{cs}$ is produced in a collision with one of the nucleons $N_1$ of the target nucleus, but not necessarily at zero angle and after scattering on another nucleon of the target $N_2$, moving at a zero angle without interaction, leaves the target nucleus.

$$\begin{array}{c} D(pn) + N_1 = \pi_{cs} + X_1 \\ \downarrow \\ \pi_{cs} + N_2 = \pi(0^o) + X_2 \end{array}. \qquad (20)$$

3. The cascade mechanism with an intermediate nucleon passes in the same way, but the intermediate particle is a nucleon $N_{cs}$, which in the second collision produces a pion emitted from the nucleus at a zero angle without interaction:

$$\begin{array}{c} D(pn) + N_1 = N_{cs} + X_1 \\ \downarrow \\ N_{cs} + N_2 = \pi(0^o) + X_2 \end{array}. \qquad (21)$$

In [9] it is shown that the cross section of the cumulative pion in the cascade process with the intermediate pion (21) does not exceed 0.6% of the cross section of the direct process in the fragmentation of the deuterium on the nucleus on the lead nucleus, and drops sharply with decrease of the target nucleus mass. In the same article it is shown that the cross section of the cumulative pion in the cascade process with the intermediate nucleon is not greater than 0.1% of the cross section of the direct process. Without details of the simulation, we note that for the twice-cumulative region the maximum contribution of the cascade process - as with the intermediate pion and with the intermediate nucleon did not exceed 0.1%. The evaluation of the cross section of twice-cumulative pions was taken from [16]. The contribution of the cascade processes was not taken into account in further calculations.

## Simulation results

We give the results of the calculation of the cross section in the deuteron fragmentation into twice-cumulative pions for the next three models of the flucton.

1. Model of the spherical flucton. In this model flucton is considered to be the configuration, when two or more nucleons of the target are situated into a spherical volume of the radius R [1,4]. Further this radius R will be called the radius of the flucton. In the interval of a flucton radii $0.4\,\text{fm} \leq R \leq 0.9\,\text{fm}$ the cross section dependence on the atomic mass of the target nucleus is the same. But in this case, the absolute values of the cross sections increase with increasing of flucton radius, changing the value by 6 times (in proportion to the cube of the radius of the flucton).
2. The other model, for which the simulation was carried out, was the model of cylindrical flucton (tube model) [17]. In this model flucton is a cylinder with the radius R and the length L along the direction of the beam particle.
3. Another model, which was used for simulation, was a model of a spherical fluctons consisting of the pair of a proton and a neutron, if they fall into a spherical volume with a radius of R (n, p - flucton). It was assumed that the two protons or two nucleons pair does not forme such kind of flucton.

The simulation results of the cross section of twice-cumulative pions production from the atomic mass of the target nucleus for the some parameters of the fluctons are shown in Fig. 4 and Fig. 5 .

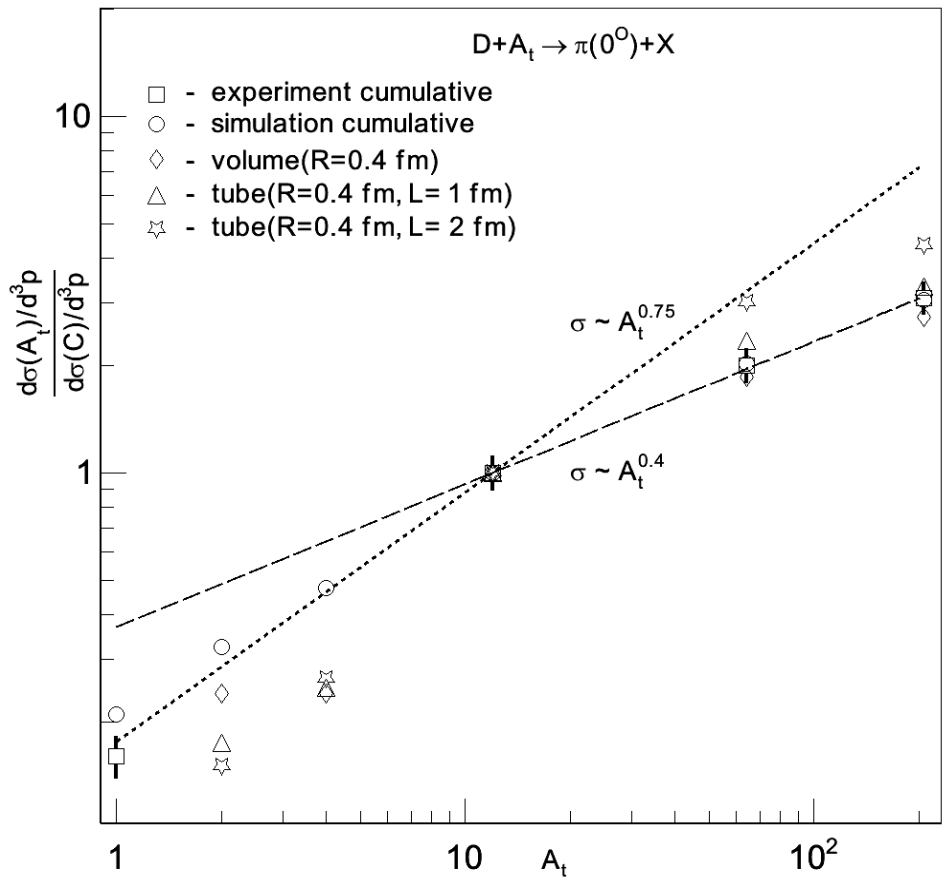

Рис. 4. Simulation results of the dependence of the cross section of deuteron fragmentation into cumulative and twice-cumulative pions (7) on the atomic mass of the target nucleus for NN spherical and cylindrical fluctons with two length parameters. Open squares - experimental data for cumulative pions from [5]. Open circles - the result of simulation for cumulative pions in the approach used in this paper (see. [9]). Lines are drawn to illustrate the nature of the atomic mass dependences.

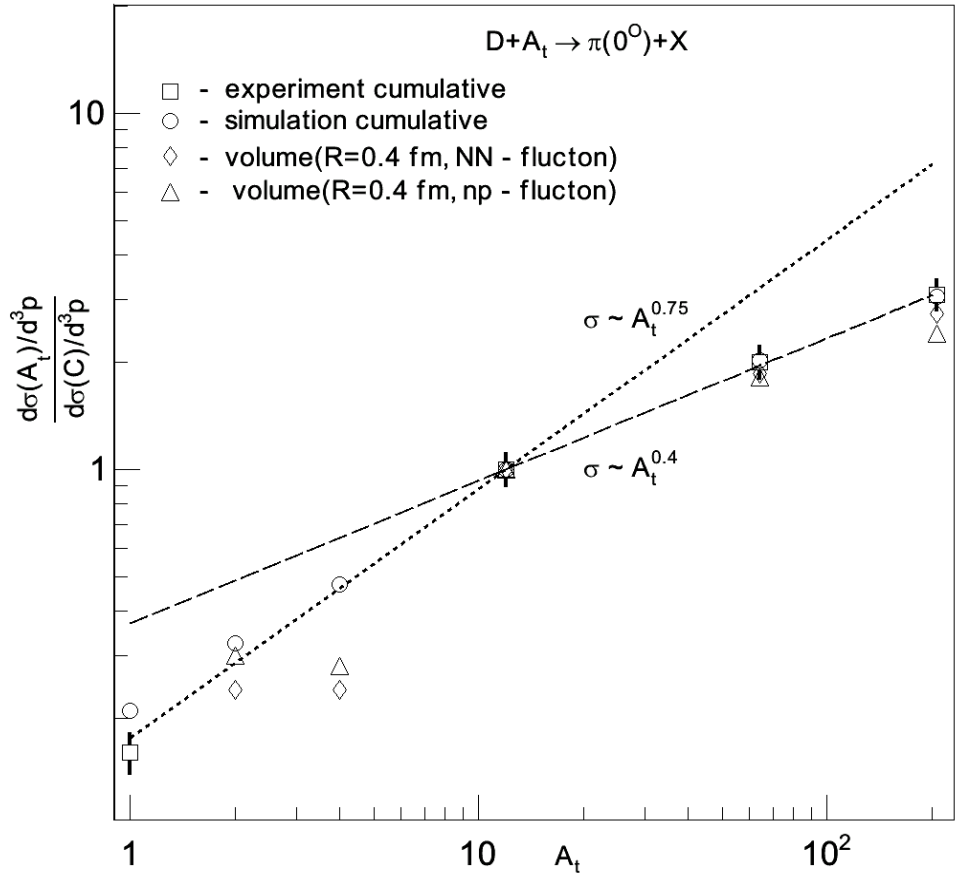

Fig. 5. Simulation results of the dependence of the cross section of deuteron fragmentation into cumulative and twice-cumulative pions (7) on the atomic mass of the target nucleus for NN spherical and np spherical fluctons. Open squares - experimental data for cumulative pions from [5]. Open circles - the result of simulation for cumulative pions in the approach used in this paper (see. [9]). Other symbols are explained. Lines are drawn to illustrate the nature of the atomic mass dependences.

## Conclusion

In this paper the reaction of the deuteron fragmentation into twice-cumulative pion was discussed. The dependence on the atomic mass of the target nucleus for pion production cross section at zero angle in twice-cumulative region was investigated with the use of simulation based on hadron-hadron scattering. This simulation has been successfully used in describing the dependence of the target nucleus cross section reaction of the incident deuteron fragmentation into cumulative pions produced at zero angle on the atomic mass. A distinctive feature of the production of twice-cumulative pions is that for their production there should be high-momentum components (fluctons) as in the deuteron and the nucleus of the target. This means that, in principle, the cross section of cumulative pion production should be sensitive to the structure of the target nucleus at small inter-nucleon distances. The simulation was performed for the three models of fluctons:
1. flucton of a spherical shape;
2. flucton of a tube shape;
3. flucton of a spherical shape consisting of a proton and a neutron.

From the simulation results it follows that for a spherical model of fluctons the dependence of the cross section production cross section on the atomic mass of the target nucleus in the cumulative and twice-cumulative region is close to each other for medium and heavy nuclei ($A \geq 12$). The difference is observed in the region of light nuclei (D, $^4$He), where the cross section production of double-cumulative pions is normalized to the carbon nucleus by 2-3 times less than similar sections for cumulative pions (see. Fig. 4). The dependence of the target nucleus cross section on the atomic mass of the target nucleus for spherical and cylinder models of fluctons leads to a different dependence on the target nucleus for light and for heavy nuclei (see. Fig. 4). A model of a spherical flucton consisting of a proton and a nucleon gives results close to the spherical fluctons consisting of two nucleons.

## Bibliography

1. A.M. Baldin, PEPAN, **V.8(3**), 429, (1977)
2. V.K. Lukyanov, A.I. Titov, PEPAN, **V.10(4)**, 815, (1979)
3. V.S. Stavinski, PEPAN, **V.10(5**), 949, (1979)
4. A.V. Efremov, PEPAN, **V.13(3**), 613, (1982)
5. A.G. Litvinenko, A.I. Malakhov, P.I. Zarubin, PEPAN Lett., V.1[58], 27, (1993)
6. Yu.S. Anisimov et al., Nucl.Phys., **V.60**, 1070, (1997)
7. V.K. Bondarev et al., PEPAN Lett., **V.4**, 4, (1984)
8. D.I. Blohintsev, JETF, **V.33**, 1295, (1957)
9. L.L. Frankfurt and M.I. Strikman, **V.76**, 215, (1981)
10. L.S. Zolin, V.F. Peresedov, PEPAN Lett., **V.3[54]**,593, (1992)
11. E.Moeller et al., Phys.Rev.C, **V.C28(3)**,1246, (1983)
12. Kh. Abraamyan et al., Phys. Lett., **V.B323**, 1, (1994)
13. A. Amaya and E. Chacon, Comp. Phys. Comm., **V.71**, (1992)
14. M. Sagavara, L. Hulthen, Handb. Phys., V.39, 1, (1957)
15. Barlet R.C., Jakson D.F., Nuclea Sizes and Structure, N.Y.: Oxford Univ.Press., (1997)
16. A.G. Litvinenko, A.I. Malakhov, P.I. Zarubin, PEPAN Lett., V.1[58], 27, (1993)
17. Berlad G., Dar A., and Eilam G., Phys.Rev., D13, 161, (1976)